\documentclass{llncs}

\usepackage{pst-tree}
\usepackage{color}
\usepackage{tabularx}
\usepackage{epsfig}
\usepackage{amsmath}
\usepackage{amssymb}
\usepackage{graphicx}
\usepackage{wasysym}
\usepackage{comment}
\usepackage{float}
\usepackage{algorithm}
\usepackage[noend]{algpseudocode}

\begin{document}

\frontmatter

\pagestyle{headings}

\mainmatter           

\title{NAE-SAT-based probabilistic membership filters}

\titlerunning{NAE-SAT filters}

\author{Chao Fang\inst{1} \and
Zheng Zhu\inst{1} \and
Helmut G.~Katzgraber\inst{1,2,3}}

\authorrunning{Chao Fang et al.}

\tocauthor{Chao Fang, Zheng Zhu, and Helmut G.~Katzgraber}

\institute{
Department of Physics and Astronomy, Texas A\&M University,\\
College Station, Texas 77843-4242, USA
\and
1QB Information Technologies (1QBit),\\
Vancouver, British Columbia, Canada V6B 4W4
\and
Santa Fe Institute,\\
1399 Hyde Park Road, Santa Fe, New Mexico 87501 USA}

\maketitle

\begin{abstract}

Probabilistic membership filters are a type of data structure designed
to quickly verify whether an element of a large data set belongs to a
subset of the data. While false negatives are not possible, false
positives are. Therefore, the main goal of any good probabilistic
membership filter is to have a small false-positive rate while being
memory efficient and fast to query. Although Bloom filters are fast to
construct, their memory efficiency is bounded by a strict theoretical
upper bound. Weaver {\em et al.}~introduced random satisfiability-based
filters that significantly improved the efficiency of the probabilistic
filters, however, at the cost of solving a complex random satisfiability
(SAT) formula when constructing the filter. Here we present an improved
SAT filter approach with a focus on reducing the filter building times,
as well as query times. Our approach is based on using not-all-equal
(NAE) SAT formulas to build the filters, solving these via a mapping to
random SAT using traditionally-fast random SAT solvers, as well as bit
packing and the reduction of the number of hash functions. Paired with
fast hardware, NAE-SAT filters could result in enterprise-size
applications.

\keywords{Satisfiability, Set Membership Filters, NAE-SAT}

\end{abstract}

\section{Introduction}
\label{sec:intro}

The {\em set membership problem} is ubiquitous. It appears in many
industrial \cite{tarkoma:12}, computer science
\cite{broder:04,tarkoma:12}, and security applications, and finds
applicability across many fields of research. The problem is simple to
pose: Given a pool of subjects $D$, and a set of interest $Y \subseteq D$,
determine if an element $x \in D$ belongs to $Y$. In real-world
applications the subset $Y$ is finite, however, it can be very large.
Therefore, determining if $x$ is a member of the subset $Y$ can be a
computationally expensive task.

A simple example is the following: Let $D$ all travelers crossing a
country's border in a year and $Y$ be a terrorist watch list. The set
membership problem is then to determine if a randomly-screened traveler
$x \in D$ is also a member of the watch list $Y \subseteq D$. For a country
with few travelers $|Y|$ crossing the border, this task is easily
accomplished by listing all members of $Y$ and testing if $x$ is one of
them.  However, for a large country where millions of travelers cross
the border each year, verifying that $x \in Y$ can be a time-consuming
task. It is therefore desirable to develop a {\em filter} that
quickly verifies if a particular traveler $x$ is on the watch list $Y$. In turn,
membership filters are not Boolean in nature. If an element $x \in D$ is
sent through a filter, if will return either {\em maybe} or {\em no}.
While {\em no} here is a definite no, {\em maybe} is interpreted as a
possible presence of $x\ in Y$. This means that membership filters have
a finite false-positive rate. However, the storage needed to store the
filter is considerably smaller than the storage needed to store the set
$Y$.  Furthermore, the query time is (ideally) faster than exhaustively
searching for $x$ in the set $Y$.  Within the traveler example, this
would mean that when travelers are screened using a probabilistic
membership filter, a query returning {\em no} means $x \not\in Y$.
However, should this not be the case, then $x$ would be sent to
secondary screening.

An ideal probabilistic set membership filter should therefore be fast,
have a small memory footprint, and a low false-positive rate.
Traditional work horses are Bloom filters \cite{bloom:70}. These are
fast and easy to implement. However, there is a information theoretical
upper bound on their memory efficiency.  More recently, Weaver {\em et
al.}~ \cite{weaver:14}, introduced random satisfiability-based
membership filters that significantly improved the efficiency.  At the
core of the filter lies the solution of a complex satisfiability (SAT)
formula  \cite{biere:09} needed to construct the actual filter.  In this
work we present a variation of Weaver {\em et al.}'s SAT filter approach
with a focus on improving the filter building, as well as query times.
Our approach is based on using not-all-equal (NAE) SAT formulas to build
the filters, as well as bit packing and reduction of the hash functions
to reduce the query times. We show that NAE-SAT filters have excellent
memory efficiency and are fast, therefore ideally suited for deployments
on large-scale applications.

\section{Preliminaries}
\label{sec:prelim}

Before outlining our implementation of SAT-based probabilistic membership
filters based on NAE-SAT, we remind the reader of traditionally-used
Bloom filters, as well as outline traditional random $k$-SAT and the
special case of not-all-equal (NAE) SAT.

\subsection{Reminder --- probabilistic Bloom filters}
\label{sec:bloom}

Probably the widest used probabilistic membership filters are Bloom
filters \cite{bloom:70}. Let $D$ be any set and $Y \subseteq D$ with $m
= |Y|$. We assume that the available memory for the filter $B_{Y}$ is
$n$ bits. Select a hash function $h$: $D \rightarrow \mathbb{Z}$ that
maps the elements in $D$ to the range [0:$n$) uniformly and randomly.
After having chosen the hash function, the bits of $B_{Y}$ must be
initialized to $0$. Then, for an element in $y \in Y$, we set the bit at
$h(y)$ to $1$, i.e., $B_{Y} [h(Y)] \equiv 1$. To store all $y \in Y$,
store all elements of $Y$ one by one.  In most implementations there are
multiple hash functions $h_{1}$, $h_{2}$, $\ldots$ $h_{k}$. In that case
$h_{1}(y)$, $h_{2}(y)$, $\ldots$ $h_{k}(y)$ must be computed first.
Once that is completed, the bits of $B_{Y}$ are all set to $1$ at the
respective locations (for a pseudo-code version of the full Bloom filter
algorithm, see, for example, Ref.~\cite{weaver:14}).

To query the filter with an element $x \in D$, simply verify that
the bits at all the locations $h_{1}(x)$, $h_{2}(x)$, $\ldots$
$h_{k}(x)$ are set to $1$. Only when the bits at those locations are
$1$, will the filter return a {\em maybe}, otherwise the filter returns
a definite {\em no}. In the latter case $x \not\in Y$.

While Bloom filters are relatively fast, there is a
information-theoretical limit to their memory efficiency
\cite{weaver:14}. Therefore, fast probabilistic membership filters with
a better memory footprint are desirable.

\subsection{Reminder --- random $k$-SAT and NAE SAT}
\label{sec:sat}

The problem of determining if there
exists an assignment of Boolean variables that satisfies a Boolean
formula, i.e., such that the Boolean formula evaluates to {\em true} is known
as the satisfiability problem (SAT). 
If the formula evaluates to {\em true} it is called {\em satisfiable},
otherwise {\em unsatisfiable}. Random $k$-SAT \cite{biere:09} is a
special case where finite-domain constraint-satisfaction problems
\cite{apt:03,dechter:03} are encoded as Boolean functions. They are
usually expressed in a conjunctive normal form (CNF) formula, i.e.,
\begin{equation}
C_{1} \wedge C_{2} \cdots \wedge C_{m} , 
\label{eq:cnf}
\end{equation}
where $\wedge$ represents conjunction (logical AND) and each $C_{i}$ ($1
\leq i \leq m)$ represents a {\em clause}, i.e., an expression of the
form
\begin{equation} 
l_{i,1} \vee \cdots \vee l_{i,k_{i}} .
\label{eq:clause}
\end{equation}
Here, $\vee$ represents a logical disjunction (logical OR) and each
$l_{r,s}$ is called a {\em literal}. Note that literals are Boolean
variables that can also appear negated. A pair of literals is said to be
complementary if both are the same variable but have opposite signs. 

An assignment is a function $f$ from the set of variables to the set
$\{0,1\}$. An assignment satisfies $x_{i}$ if $f(x_{i}) = 1$ and
satisfies $\overline{x}_{i}$ if $f(x_{i})$ = 0. If at least one of the
literals in a clause is satisfied, we say that this clause is satisfied.
Only if all the clauses in the CNF are satisfied, the CNF is said to be
{\em satisfied}.

If there are $k$ different literals (not including complementary pairs)
in every clause, then the CNF is called a $k$-SAT instance.  In random
$k$-SAT instances, the clause-to-variable ratio $\alpha = m/n$ (where
$m$ is the number of clauses $C_{i}$ and $n$ the number of variables)
plays an important role.  Interestingly, there is a complexity phase
transition in the solvability of an instance as a function of $\alpha$.
For small $\alpha$, satisfied instances are easy to find. For large
$\alpha$ instances can almost never be satisfied and for a critical
$\alpha = \alpha_c$ a phase transition \cite{yeomans:92} occurs. For $k
< 3$, random $k$-SAT formulae can be solved in polynomial time and, for
example, $\alpha_c(k=2) = 1$.  However, for $k \geq 3$ random $k$-SAT
problems fall into the NP-complete complexity class.  This means there
are no polynomial-time algorithms (to date) that can solve these. In
particular, $\alpha_c(k = 3) \approx 4.17$ \cite{nishimori:01}.

For regular $3$-SAT, if an assignment satisfies all the literals in a
clause, the clause is considered as satisfied. The special not-all-equal
(NAE) $3$-SAT case, however, requires at least one literal to be true
and at least one literal to be false.  Therefore, the case where all
three literals are true is not allowed.

\subsection{Reminder --- SAT filters}
\label{sec:satfilter}

There are two ways to build a SAT filter. Here we only discuss the
single-instance filter. For details on how to build a filter with more
than one instance see Ref.~\cite{weaver:14}. The following steps are
needed to build a probabilistic membership filter based on SAT formulas:

\begin{itemize}

\item[$\RHD$]{{\bf Build a CNF} --- We use a set of hash functions
$h_{1}$, $h_{2}$, \ldots, $h_{k}$ to create a random clause $C_{Y}$ with
$k$ literals for each $y \in Y$ with $|Y| = m$. This means there are
$m$ clauses, which make a random $k$-SAT instance (CNF) $X_{Y}$. During
the process, it is important to ensure that all literals are different
in each clause. Furthermore, the clause-to-variable ratio $m/n$ should
not be too high such that the CNF is not in the unsatisfied regime.}

\medskip

\item[$\RHD$]{{\bf Find solutions to the CNF} -- Once the CNF has been
constructed, multiple uncorrelated solutions are needed to construct a
good filter. It is important to use SAT formulas for which efficient
solvers are known to speed up this step of the filter building process.}

\medskip

\item[$\RHD$]{{\bf Store the filter} --- After finding a number of
(ideally uncorrelated) solutions, these are stored in an array as the
filter. Note that the storage requirements for the filter are
considerably smaller than the original data.}

\end{itemize}
Note that SAT filters do not allow insertions after they have been
built, i.e., adding a new element to $D$ will require a new filter to be
constructed.

To query the filter, the first step is to generate a clause $C_x$ from
the elements $x \in D$ using the hash functions used to generate the
random clause in the first step of the filter construction process. Then
one has to verify if $C_{x}$ is satisfied by the filter, i.e., {\em all}
solutions stored.  If so, the filter returns \emph{maybe}. However, if
$C_{x}$ is not satisfied by any one of the solutions, the filter returns
a definite {\em no}.  A filter is characterized by its false positive
rate, which should be as low as possible, its memory efficiency, as well
as ideally short build and query times. We discuss both for the case of
SAT filters in what follows.

\subsubsection{False positive rate ---}
\label{sec:fpr}

Probabilistic membership filters have a finite false positive rate
(FPR). This means that for an element $x$ that is not in the set of
interest $Y$ the filter might still return a {\em maybe} result.  The
FPR for SAT filters is equivalent to the probability that a random
$k$-SAT clause can be satisfied by a specific solution. For a random
$k$-SAT clause, the probability that the clause can definitely not pass
one regular solution is $2^{-k}$. Therefore, for a single solution to
the CNF, the FPR is $(1 - 2^{-k})$. This can be improved, by using
$s > 1$ solutions to the CNF, i.e., for $s$ solutions
\begin{equation}
\label{eq:fprreg}
p_{\rm{SAT}} = (1-2^{-k})^{s} .
\end{equation}
Using $s > 1$ solutions reduces the FPR, but increases both query times
and storage requirements by a factor $s$. Note that if the solutions are correlated,
then the FPR might not be reduced by increasing the number of solutions.
Therefore, it is imperative to use a SAT solver that produces as uncorrelated
solutions as possible (i.e., with a large hamming distance).
We note that for the special case of building SAT filters with NAE-SAT
formulas Eq.~\eqref{eq:fprreg} changes to
\begin{equation}
\label{eq:fprnae}
p_{\rm{NAE}} = (1-2^{-k+1})^{s} .
\end{equation}

\subsubsection{Efficiency ---}
\label{sec:efficiency}

The memory efficiency of a probabilistic membership filter is defined
as the number of filter bits required per keyword item. As introduced
in Ref.~\cite{weaver:14}, the memory efficiency $\xi$ for a SAT filter
is given by
\begin{equation}
\emph{$\xi$} = \frac{-\log_2 p}{n/m} .
\end{equation}
Here, $n$ is the number of memory bits and $m$ = $|Y|$. For a SAT filter
that uses $s$ solutions, one needs $sn$ memory bits and therefore the
efficiency is given by
\begin{equation}
\label{eq:efficiencyreg}
\xi_{\rm{SAT}} = \frac{-\log_2 p_{\rm{SAT}}}{sn/m} = \frac{-\log_2
(1-2^{-k})}{n/m} .
\end{equation}
Again, for the special case that uses NAE-SAT solutions,
Eq.~\eqref{eq:efficiencyreg} changes to
\begin{equation}
\label{eq:efficiencynae}
\xi_{\rm{NAE}} = \frac{-\log_2 (1-2^{-k+1})}{n/m} .
\end{equation}
It can be shown rigorously \cite{weaver:14} that SAT filters can achieve
a theoretical efficiency of $1$ (i.e., 100\%). This is to be contrasted
to Bloom filters that have an information-theoretical upper bound for
the efficiency, namely $\xi_{\rm Bloom} \le \log 2 \approx 0.698$.
This is the main reason why SAT filters are more desirable than Bloom
filters.

\subsubsection{Build \& query times ---}
\label{sec:queryefficiency}

In Ref.~\cite{weaver:14} it was shown that while query times for SAT
filters are short, they are still larger than for Bloom filters.
Furthermore, the construction of a SAT filter requires multiple
uncorrelated solutions to a CNF. While efficient SAT solvers exist,
some tend to produce correlated solutions, which thus means that there
is potentially a large overhead in finding as uncorrelated solutions
(i.e., with a large hamming distance) as possible.

In what follows we demonstrate how filters designed using NAE-SAT
formulas have considerably shorter build, as well as query times.
Furthermore, we show how a NAE-SAT CNF can be reformulated into a SAT
formula such that one can take advantage of state-of-the-art random
$k$-SAT solvers. Finally, we show that only one hash function is needed,
thus optimally speeding up build and query times.

\section{Building a NAE-SAT filter}
\label{sec:building}

A low FPR is of utmost importance for probabilistic membership filters.
In the case of SAT filters, the theoretical FPR [Eq.~\eqref{eq:fprreg}]
can only be reached if the solutions to the underlying SAT formula are
{\em uncorrelated}. Because the solutions are from a {\em single} SAT
formula, the probability that these are correlated is high, unless a
typically large effort to find enough uncorrelated solutions is
performed. 

Here we work around this bottleneck by replacing standard SAT formulas
with NAE-SAT formulas, because the solutions to NAE-SAT problems have
large Hamming distances (typically around $50$\% of the number of
variables) and are therefore far less correlated \cite{coja:13}. In our
approach we randomly select the NAE-SAT solutions to build the filter.
Because we never know which solution in the pair was selected, we can
state statistically that the {\em average} hamming distances would be
around 50$\%$ of the number of variables by construction.  This saves
considerable time when building the probabilistic filter. From now on,
unless otherwise specified, we use NAE-SAT solutions.

As is the case for traditional SAT filters, it desirable to have a small
value of $k$ (number of literals per clause) to reduce query times. As
we shall demonstrate, near-perfect efficiencies $\xi$ can be obtained
for $k \lesssim 6$. The number of variables in the SAT instance $n$
should be chosen such that $\alpha = m/n$ is in a regime where it is not
too hard to solve the CNF with a given SAT solver. Because the size of
the data set is given ($m = |Y|$, $m$ the number of clauses per
instance), this shall influence the choice of $n = m/\alpha$. Finally,
because there is as tradeoff between the FPR $p$ and the memory
efficiency, this is what dictates the value of $s$. In particular, the
lower the amount of required storage, the higher the FPR. From
Eq.~\eqref{eq:efficiencynae} one can estimate the number of SAT
instances $s$ needed to achieve a given FPR $p$.

Furthermore, we experiment with a SAT solver \cite{zhu:16y} that
efficiently traverses the solution space, thus generating typically
uncorrelated solutions. {\em borealis} --- a method that works extremely
well to solve both weighted and unweighted MAX-SAT problems --- is based
on parallel tempering Monte Carlo, a standard workhorse in the study of
frustrated magnetic systems in statistical physics. The idea is to
randomly propose variable changes using a simple Monte Carlo method. In
addition to the local updates, the system is replicated at multiple
temperatures \cite{hukushima:96,katzgraber:06a,earl:05}. Swaps between
temperatures are allowed, therefore allowing the system to relax out of
local minima and more efficiently sample the solution space.  {\em
borealis} is typically not faster than highly-tuned SAT solvers.
However, it is a generic method that works relatively well for many
SAT-type problems and can produce easily uncorrelated solutions.

We first analyze the performance of NAE-SAT filters using traditional
random $k$-SAT solvers and then show results using {\em borealis}. Note
that we use similar parameters as used in Ref.~\cite{weaver:14} to be
able to perform a direct comparison between the results of
Ref.~\cite{weaver:14} on traditional SAT filters and our NAE-SAT
implementation shown here.

\subsection{NAE-approach using traditional solvers}
\label{sec:building-traditional}

To be able to use traditional $k$-SAT solvers such as Dimetheus
\cite{dimetheus:14} ($\alpha \approx \alpha_c$) or WalkSAT{\em lm}
\cite{cai:13} ($\alpha \lesssim \alpha_c$) for NAE-SAT instances, we
have modified the original NAE-SAT CNF into a SAT CNF such that by
construction all solutions satisfy the NAE requirement. This is
accomplished by adding a penalty clause to rule out the all-satisfied
assignments to each clause in the original CNF. In the penalty clause
all literals are complementary to the original clause.  Then, as long as
the solver finds a solution, the solution is a NAE solution to the
original CNF. For example,
\begin{equation}
(x_{3}\vee x_{18}\vee \overline{x_{12}}\vee x_{5})
\rightarrow
(x_{3}\vee x_{18}\vee \overline{x_{12}}\vee x_{5})\wedge
(\overline{x_{3}}\vee \overline{x_{18}} \vee x_{12}\vee
\overline{x_{5}}) .
\end{equation}
The term on the right now satisfies the NAE constraint and can be
handled by traditional random $k$-SAT solvers.

\subsubsection{Build times ---}

We first benchmark the build time of a NAE-SAT filter. As an example, we
construct a filter with 75$\%$ memory efficiency and $2^{14}$ clauses.
The results are shown in Table \ref{tab:table_1}. Because $\alpha$ is
small in this case, we use Walksat\emph{lm} for the case study. In Table
\ref{tab:table_2} we quote for comparison the results of
Ref.~\cite{weaver:14} for a regular $k$-SAT filter.  The use of NAE-SAT
solutions clearly reduces the filter build times in comparison to the
$k$-SAT version of the filters. Although the used hardware is different,
the difference cannot account for the large difference in filter build
times.

\begin{table}[h]
\caption{
Build time in seconds, memory size in bytes and false-positive rate (FPR) in
percent for the $k$-NAE-SAT filter case study using Walksat\emph{lm}. By design,
the average Hamming distance is around $50$\%. Simulations were performed
on a 2013 MacBook Pro with a $2.60$ GHz processor. $s$ represents the number
of NAE-SAT instances used to build the filter.
\label{tab:table_1}
}
\begin{center}
\begin{tabular*}{35em}{@{\extracolsep{\fill}}l l l r}
\hline
\hline
Filter size & Build time (s) & Size (bytes)  & FPR (\%)\\
\hline
$k=4$, $s=11$ & $0.6$ & $44748$ & $23.00\%$\\
$k=5$, $s=22$ & $1.0$ & $44000$ & $24.45\%$\\
$k=6$, $s=44$ & $1.2$ & $44144$ & $25.10\%$\\
\hline
\hline
\end{tabular*}
\end{center}
\end{table}

\begin{table}[h]
\caption{
Build time in seconds, memory size in bytes and false-positive rate
(FPR) in percent for the $k$-SAT filter studied in
Ref.~\cite{weaver:14}.  The average hamming distance is at least 
$50$\%. Simulations were performed on a 2009 MacBook Pro with a $3.06$
GHz processor. $s$ represents the number of SAT instances used to build
the filter.
\label{tab:table_2}
}
\begin{center}
\begin{tabular*}{35em}{@{\extracolsep{\fill}}l l l r }
\hline
\hline
Filter size & Build time (s) & Size (bytes)  & FPR (\%)\\
\hline
$k=4$, $s=22$ & 20802 & 44748 & 24.20$\%$\\
$k=5$, $s=44$ & 610 & 44000 & 24.86$\%$\\
$k=6$, $s=88$ & 643 & 44144 & 25.09$\%$\\
\hline
\hline
\end{tabular*}
\end{center}
\end{table}

\subsubsection{Efficiency vs False-positive rate ---}

\begin{figure}[h]
\begin{center}
\includegraphics[width=1.00\columnwidth]{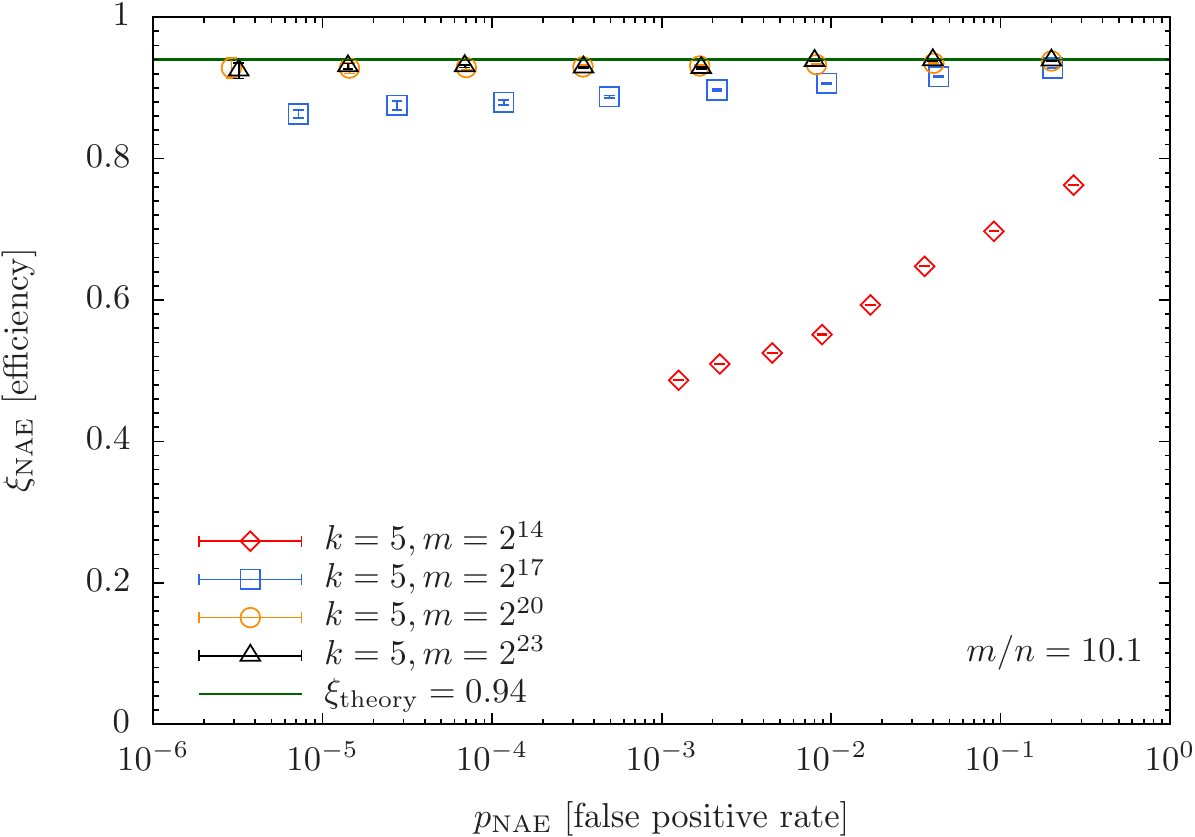}
\caption{Efficiency $\xi_{\rm NAE}$ as a function of false-positive rate
$p_{\rm NAE}$ for a $k$-NAE-SAT filter. As $m$ increases for fixed $k$,
the efficiency approaches the theoretical bound (solid horizontal line).
Note that very low FPRs can be obtained. Instances generated with
Dimetheus.
\label{fig:effi2}}
\end{center}
\end{figure}

Figure \ref{fig:effi2} shows the memory efficiency $\xi_{\rm NAE}$ vs
the FPR $p_{\rm NAE}$. Using formulas with $m/n = 10.1$ and increasing
the filter size $m$ eventually has little effect on the efficiency.
However, for increasing $m$ the FPR $p_{\rm NAE}$ can be reduced to
arbitrarily-low values; here below $10^{-5}$. The solid horizontal
(green) line represents the optimal bound which can easily be achieved
with little numerical effort. Note that the instances were generated
using Dimetheus \cite{dimetheus:14} because the ratio is close to the
threshold.

\begin{table}[h]
\caption{
Build time in seconds, memory size in bytes and false-positive rate
(FPR) in percent for the $k$-NAE-SAT filter case study using {\em
borealis}. By design, the average Hamming distance is around $50$\%.
Simulations were performed on a 2013 MacBook Pro with a $2.60$ GHz
processor. $s$ represents the number of NAE-SAT instances used to build
the filter.
\label{tab:table_3}
}
\begin{center}
\begin{tabular*}{35em}{@{\extracolsep{\fill}}l l l r }\\
\hline
\hline
Filter size & Build time(s) & Size(bytes)  & FPR\\
\hline
$k=4$, $s=11$ & 6.2 & 44748 & 23.00$\%$\\
$k=5$, $s=22$ & 11.0 & 44000 & 24.45$\%$\\
$k=6$, $s=44$ & 17.8 & 44144 & 25.10$\%$\\
\hline
\hline
\end{tabular*}
\end{center}
\end{table}

\subsection{Using physics-based solvers}
\label{sec:building-pt}

Because {\em borealis} \cite{zhu:16y} is designed to tackle statistical
physics problems, we need to cast the CNF of the NAE-SAT formula as a
physical Hamiltonian (cost function). We use the number of unsatisfied
clauses as a simple cost function for the $n$ Boolean variables in the
CNF. If a cost of $0$ is found (in physics, the ground state energy),
the configuration represents a valid variable assignment to the CNF.
Details on the algorithm and its implementation can be found in
Ref.~\cite{zhu:16y}.

\subsubsection{Build times ---}

Table \ref{tab:table_3} lists our results for the same instances studied
in Tab.~\ref{tab:table_1}. Although {\em borealis} is slower than
Dimetheus for these instances, using NAE-SAT filters is still
considerably faster than $k$-SAT filters (see Tab.~\ref{tab:table_2}.
We do note, however, that {\em borealis} works reasonably well for a
broad range of $\alpha$ values unlike traditional SAT solvers that are
tuned to specific regimes of $\alpha$ values.

\subsubsection{Efficiency vs false-positive rate ---}

\begin{figure}[h]
\begin{center}
\includegraphics[width=1.00\columnwidth]{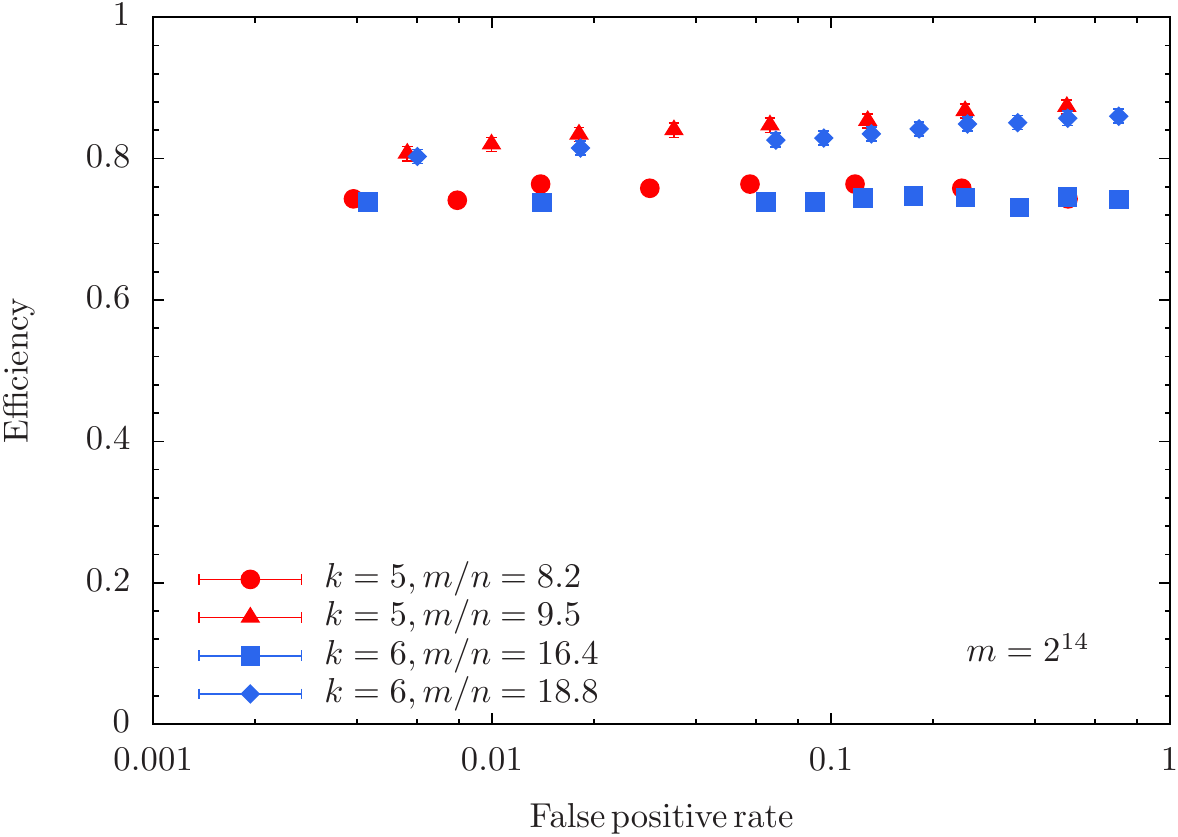}
\caption{Efficiency $\xi_{\rm NAE}$ as a function of false-positive rate
$p_{\rm NAE}$ for a $k$-NAE-SAT filter. Data computed with {\em
borealis} for different values of $\alpha = m/n$. Finite-size effects
(see text) are visible due to the small filter size used. However, {\em
borealis} is an efficient solver for a broad range of $\alpha$ values.
\label{fig:fpr_effi_1}}
\end{center}
\end{figure}

Figure \ref{fig:fpr_effi_1} shows the efficiency $\xi$ vs FPR $p$ for a
fixed filter size $m = 2^{14}$ and different values of $\alpha = m/n$
($k = 5$). {\em borealis} performs reasonably well for a broad range of
$\alpha$ values with $ 8 \lesssim \alpha \lesssim 19 < \alpha_c \approx
21.11$. There is a decrease of the efficiency $\xi$ for small FPRs. This
is because $m$ is relatively small and therefore the number of
uncorrelated solutions is accordingly small. As such, finding a set $s$
of many uncorrelated solutions is difficult. This problem is referred to
as {\em finite-size effect} in physics and is easily alleviated by
increasing the filter size. For $n/s$ large the finite-size effects
become negligible.  We expect that a better implementation of {\em
borealis} might show a clear advantage over traditional solvers for
larger filter sizes. We leave this study for a future publication.

\subsection{Querying NAE-SAT filters}
\label{sec:querying}

To speed up query times for the NAE-SAT filter we use bit packing. Given
the 64-bit architecture of the benchmark machine, this means that 64
instances can be handled in parallel.

\begin{figure}[h]
\begin{center}
\includegraphics[width=1.00\columnwidth]{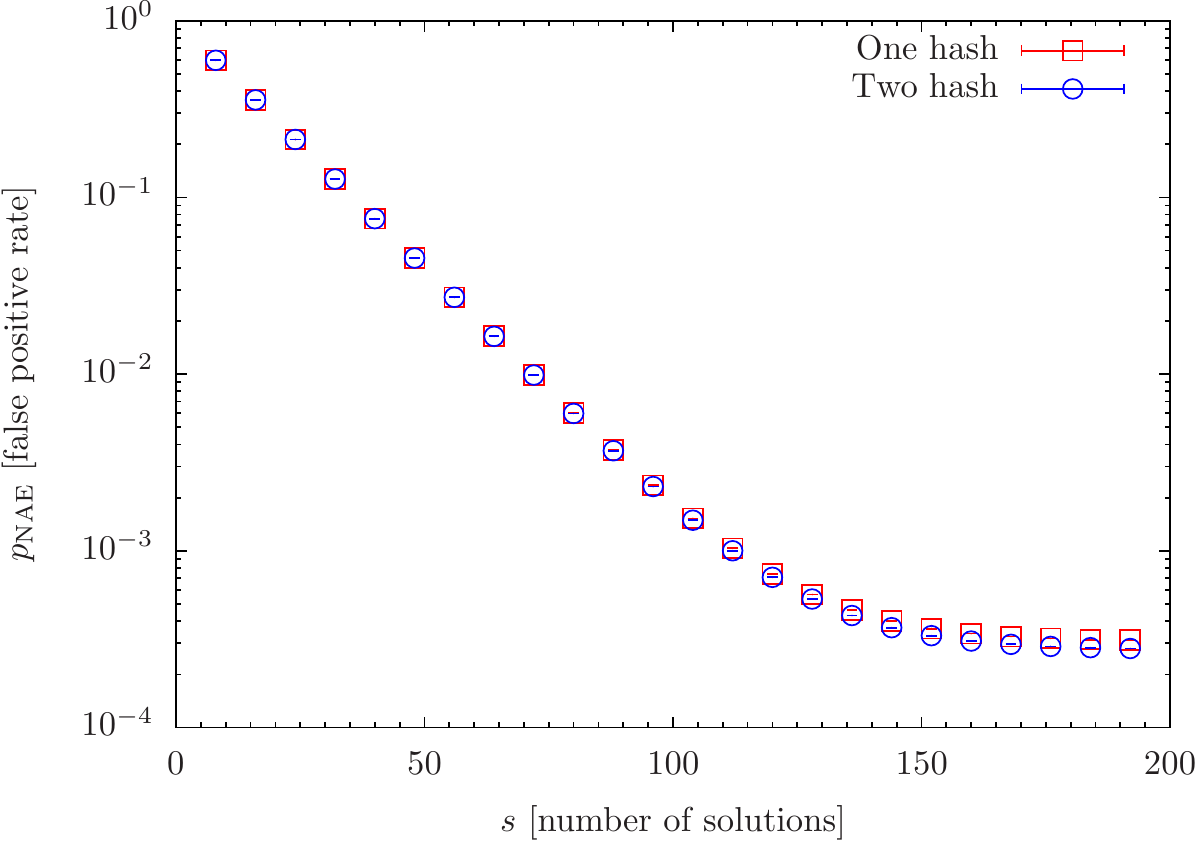}
\caption{FPR $p$ as a function of the number of solutions $s$ using one, or
two-hash functions in \texttt{MurmurHash3} \cite{murmurhash3:11}. The
difference between both approaches is $\sim 10^{-5}$, i.e., negligible.
\label{fig:comp}}
\end{center}
\end{figure}

For the benchmarks, we use a single core 2013 MacBook Pro with a
$2.60$GHz processor and 8Gb RAM. We query $2^{17}$ 64-bit strings. The
hash function used is \texttt{MurmurHash3} \cite{murmurhash3:11}.  Note
that we deviate from the approach used in Ref.~\cite{weaver:14},
because, as Fig.~\ref{fig:comp} shows, the difference between the FPR
using one or two hashes is negligible. We did simply change the seed in
\texttt{MurmurHash3} to achieve these results. This is yest another
advantage of our implementation that speeds up query times.

\begin{figure}[h]
\begin{center}
\includegraphics[width=1.00\columnwidth]{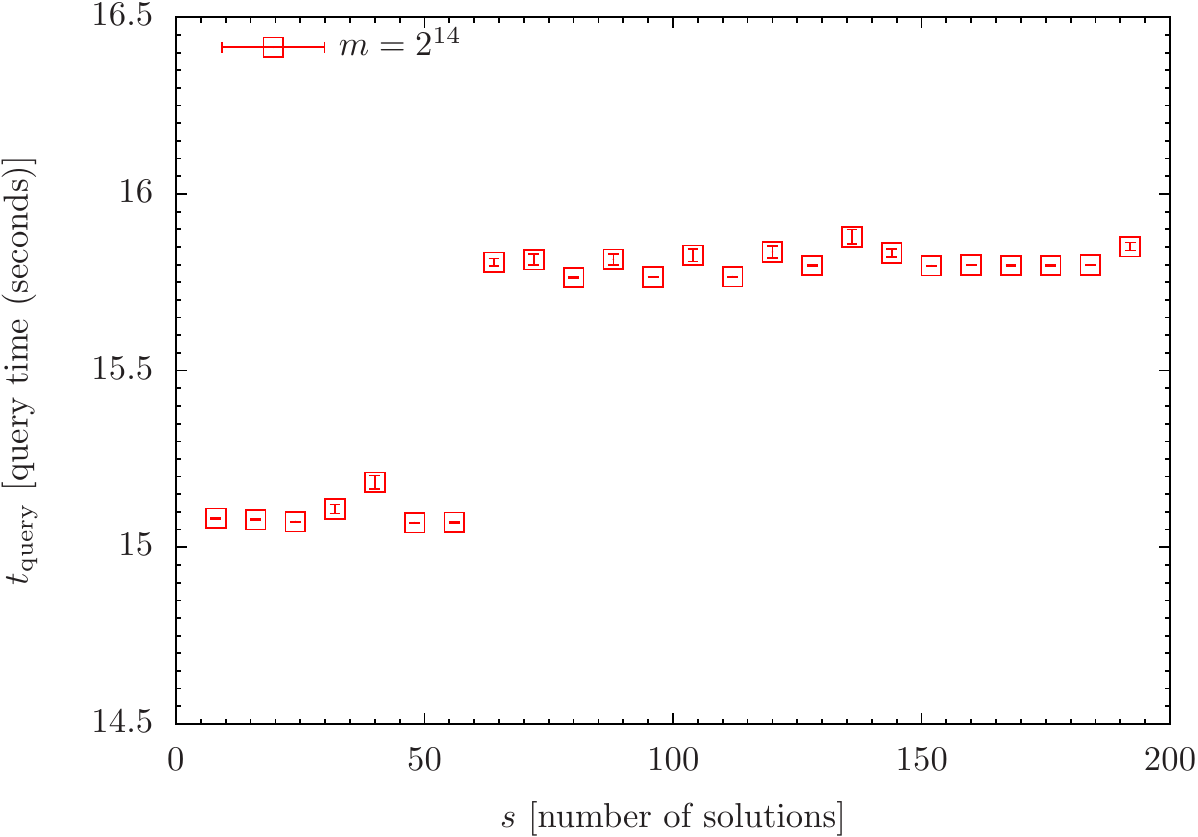}
\caption{Query time as a function of the number of used instances $s$ for a
NAE-SAT filter with $k = 5$, $m = 2^{17}$, and $n = 16282$. Queried 
are $2^{17}$ 64-bit strings. Time $t$ is measured in seconds.
\label{fig:query}}
\end{center}
\end{figure}

Figure \ref{fig:query} shows that the query times are approximately
similar for different values of $s$, as long as there are more than a 
certain number of solutions. The jump in the data might be due to 
buffering issues in the bit packing.

\section{Summary}
\label{sec:conclusion}

We have demonstrated that by using NAE-SAT problems for the construction
of SAT filters as first mentioned in Ref.~\cite{weaver:14}, filter build
times can be considerably reduced. Using NAE-SAT formulas to build the
filters has the advantage that, by design, the solutions tend to be
uncorrelated. Furthermore, we show how the NAE-SAT constraint can be
accommodated into a random SAT formula such that standard SAT solvers can
be used. Query times in our implementation are reduced via bit packing
and the use of a single hash function.  Finally, by using
physics-inspired algorithms such as {\em borealis} the filter
construction can be further parallelized and improved further because
the algorithm efficiently traverses the solution space.

\section*{Acknowledgments}

We would like to thank Bryan Jacobs, Brad Lackey, Oliver Melchert, and
Andrew Ochoa, for fruitful discussions. H.G.K.~would like to thank
Diplomatico Reserva for inspiration. The authors acknowledge support
from the National Science Foundation (Grant No.~DMR-1151387).  The
research is based upon work supported by the Office of the Director of
National Intelligence (ODNI), Intelligence Advanced Research Projects
Activity (IARPA), via Interagency Umbrella Agreement IA1-1198, as well
as via MIT Lincoln Laboratory Air Force Contract No.~FA8721-05-C-0002.
The views and conclusions contained herein are those of the authors and
should not be interpreted as necessarily representing the official
policies or endorsements, either expressed or implied, of the ODNI,
IARPA, or the U.S.~Government. The U.S.~Government is authorized to
reproduce and distribute reprints for Governmental purposes
notwithstanding any copyright annotation thereon.  We thank Texas A\&M
University and the Texas Advanced Computing Center (TACC) at The
University of Texas at Austin for providing HPC resources.

\bibliography{refs,comments}

\end{document}